\documentclass[12pt,epsfig,rotate]{article}
\usepackage{epsfig}
\usepackage{wrapfig}
\usepackage{graphicx}
\begin{document}

\setcounter{page}{1}

\begin{center}
{\bf
QUASI-ELASTIC SCATTERING  OF  PROTON WITH 1 GEV  ENERGY
 ON EIGHT-NUCLEON CLUSTER INSIDE NUCLEUS}\\

\vspace*{9mm}
{\bf O.V.~Miklukho, A.Yu.~Kisselev, G.M.~Amalsky, V.A.~Andreev,
G.V.~Fedotov, G.E.~Gavrilov, A.A.~Izotov,  N.G.~Kozlenko, P.V.~Kravchenko, 
V.I.~Murzin, D.V.~Novinskiy, A.V.~Shvedchikov, V.A.~Stepanov, and A.A.~Zhdanov}\\
\vspace*{5mm}
{\it B.P.~Konstantinov Petersburg Nuclear Physics Institute, National Research Centre
Kurchatov Institute, 
Gatchina, 188300 Russia}\\
\end{center}

\vspace*{10mm}
Available data on the polarization of the secondary proton
(as a function of its momentum $K$) in the inelastic ({\it p, p$'$}) reactions with
the $^{9}$Be, $^{12}$C, and $^{40}$Ca nuclei and differential cross section data (the momentum distributions) for the reactions 
at the initial proton energy 1 GeV and scattering angles $\Theta$=21$^\circ$ and $\Theta$=24.5$^\circ$  
were analysed in a range of the high momenta  $K$ close to the momentum corresponding to the proton elastic scattering off the investigated nucleus.  
A structure in the polarization
and momentum distribution data at the high momenum $K$, related probably to quasi-elastic scattering off a $^{8}$Be-like nucleon cluster
inside the nuclei, was observed.\\
\\
PACS numbers: 13.85.Hd, 24.70.+s, 25.40.Ep, 29.30.-h\\
\\
\\
{\bf Comments:} 14~pages, 11~figures. \\
\\
\\
\\
{\bf Category:} Nuclear Experiment (nucl--ex)\\

\newpage
\section{Introduction}

 This  work is a part of the experimental program in the framework of which
the effects of nucleon clusterization in nuclear matter is
studied in  the inclusive  ({\it p,~p$'$})  experiments 
at the PNPI synchrocyclotron with the 1~GeV proton beam 
\cite{Miklukho2011, Miklukha2015, Miklukho2017, Miklukho2018, Miklukho2020,   Miklukha2017, MiklukhaC2017}.

The general layout of the experimental setup  is presented in Fig.~1. 

\begin{figure}
\centering\epsfig{file=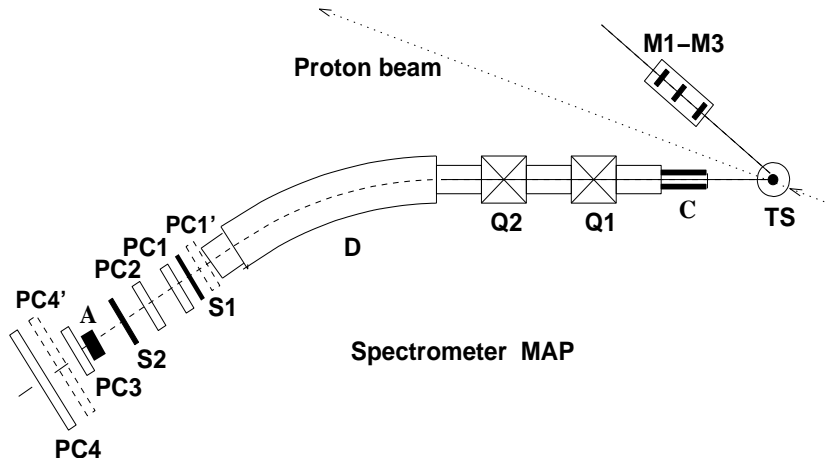,width=.75\textwidth=100mm,height=60mm}
\caption{\small The experimental setup. TS is the target of the MAP spectrometer;
Q1$\div$Q2 are the magnetic quadrupoles; D is the dipole
magnet; C1 is the collimator; S1$\div$S2 and M1$\div$M3 are
the scintillation counters; PC1$\div$PC4, PC1$'$, PC4$'$
and A are the proportional
chambers and the carbon analyzer of the MAP
polarimeter, respectively.}
\end{figure}
The proton beam was focused
onto the  target TS of the magnetic  spectrometer MAP. The beam intensity 
was monitored by the scintillation telescope M1, M2, M3. 
The spectrometer was used to measure the momenta of the secondary protons  
from the inclusive  ({\it p,~p$'$}) reaction as well as their polarization.
The momentum of the scattered proton ($K$) was determined using the coordinate information 
from the proportional chamber  PC2-X. The momentum resolution of the spectrometer
($\pm$ 2.5 MeV/c)
was estimated by measuring the width of
the clearly separated 2$^+$ excited level in the ({\it p, p$'$}) reaction 
with the $^{12}$C nucleus at the scattering angle 21$^\circ$ under investigation (Fig.~2) \cite{Miklukho2017}.
In this measurement  we 
also observed a peak  (Fig. 2) which was first identified as the 1$^+$ excited level predicted 
in \cite{Viollier1975}.
The polarization of secondary protons in the ({\it p, p$'$}) reaction  was found from an azimuthal asymmetry of the proton scattering
off the carbon analyzer A, using the track information from the proportional chambers 
(PC1$\div$P4 and PC1$'$, PC4$'$) of the polarimeter.
 The main parameters of the MAP spectrometer and the
polarimeter are given in \cite{Miklukho2017}.
\begin{figure}       
\centering\epsfig{file=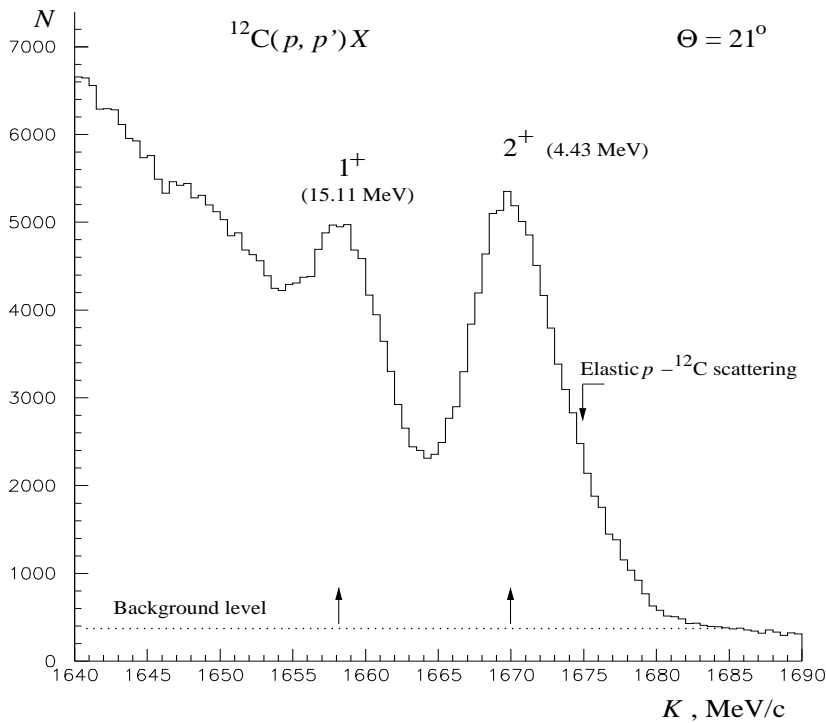,width=.75\textwidth,height=95mm}
\caption{\small Momentum distribution in the inclusive reaction $^{12}$C($p, p'$)X  at a 
 scattering angle $\Theta$=21$^\circ$ \cite{Miklukho2017}. 
 }
\end{figure}

Earlier,
 the secondary proton polarization ($P$) and its momentum distribution (differential cross section 
 $\sigma^{incl}$~=~$\frac{d^2\sigma}{d\Omega dK}$) 
in the inelastic ({\it p,~p$'$})  reaction with nucleus
were measured at the scattered angle of  $\Theta$=21$^\circ$.
The nuclei  $^{9}$Be \cite{Miklukho2020},  $^{12}$C  \cite{Miklukho2017} , $^{28}$Si  \cite{Miklukho2018},  $^{40}$Ca  \cite{Miklukho2017}, $^{56}$Fe  \cite{Miklukho2018}, 
and  $^{90}$Zr  \cite{Miklukho2020} were investigated.
The structure in these experimental  data was observed that possibly related to quasi-elastic scattering
in the momentum intervals II, III, and IV 
on two-nucleon ($^2$H), three-nucleon ($^3$He,$^3$H), and  four-nucleon ($^4$He) correlations, respectively \cite{Blokhintsev1957, CLAS2006}.

Recently the momentum distributions of the secondary protons in the  ({\it p, p$'$}) reaction with nuclei
$^{9}$Be, $^{12}$C, and $^{40}$Ca  were measured at the scattering angle of  $\Theta$=24.5$^\circ$.
The measurements were done
in the range of  high momenta of the scattered proton  only.
 
\section{Analysis of experimental data}


\begin{figure}
\centering\epsfig{file=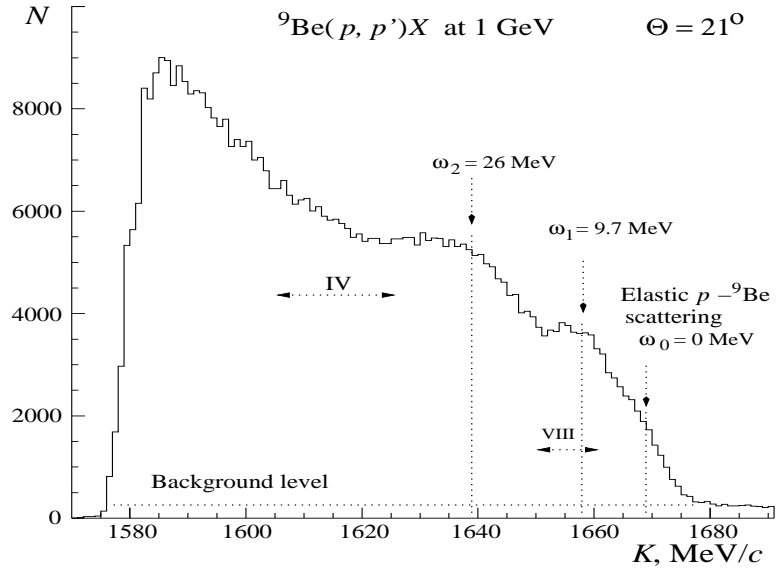,width=.70\textwidth,height=75mm}
\caption{\small Momentum distribution of the protons scattered  at an angle $\Theta$~=~21$^\circ$ 
 in the inclusive reaction $^{9}$Be($p, p'$)X. $K$ is the
secondary proton momentum. 
 A difference of the secondary proton energy calculated for the elastic proton scattering off the nucleus
 under investigation and measured in the experiment
 is signified by $\omega$. 
 The momentum interval VIII (IV) corresponds to  quasi-elastic scattering 
 on a $^{8}$Be - like (a $^{4}$He - like)  nucleon cluster.
}
\end{figure}
%
\begin{figure}
\centering\epsfig{file=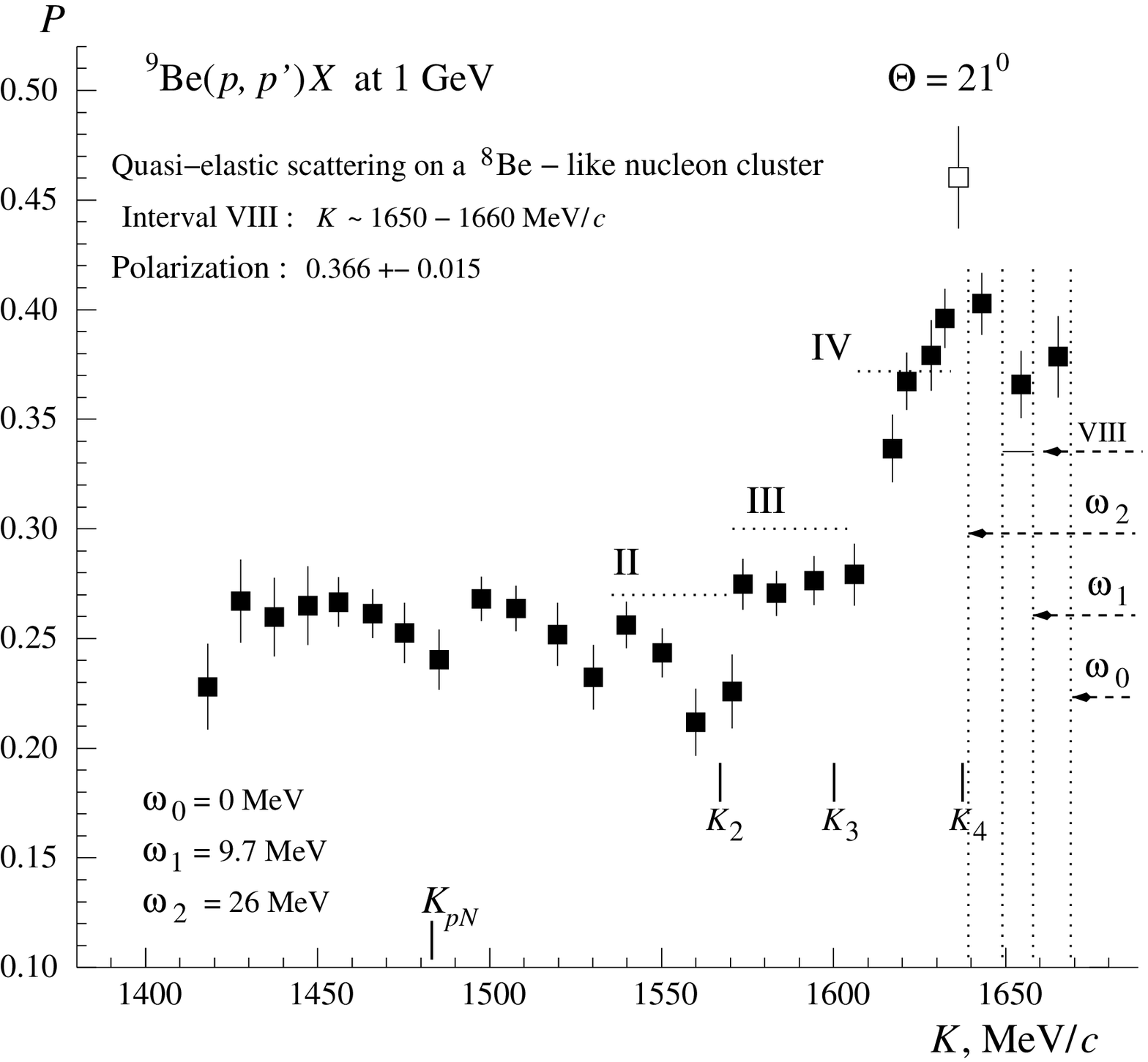,width=.95\textwidth,height=115mm}
\caption{\small  Polarization $P$ of the protons scattered  at an angle $\Theta$~=~21$^\circ$
(black squares) in the inclusive reaction $^{9}$Be($p, p'$)$X$ versus the
secondary proton momentum $K$. 
The empty square corresponds to the polarization
in the elastic $p$~-~$^4$He scattering \cite{Miklukho2006}. The dotted lines cover
the $K$ intervals II, III, and IV corresponding to quasi-elastic scattering on two-nucleon ($^2$H), three-nucleon ($^3$He, $^3$H),
and four-nucleon ($^4$He) correlations, respectively. The calculated secondary proton momenta 
for the maxima of the quasi-elastic peaks in the $^9$Be($p, p'$~NC)$X$ reaction with the corresponding nucleon correlation (NC) are designated as
 $K_2$, $K_3$, and $K_4$. A narrow  momentum interval VIII corresponds to the  quasi-elastic proton scattering on a $^8$Be-like nucleon cluster.
 Momentum $K_{pN}$ corresponds to the maximum of the quasi-elastic peak in the proton scattering off nuclear nucleons.
}
\end{figure} 
 The measured momentum distribution of the secondary proton
produced  in the reaction $^{9}$Be($p, p'$)X   at the angle of $\Theta$~=~21$^\circ$ with the momentum $K$ 
and its polarization are presented in Fig.~3 and Fig.~4, respectively \cite{Miklukho2020}.

In Fig.~3, there is a wide peak in the momentum distribution with energy $\omega$ = $\omega_2$ = 26 MeV
transferred into the nucleus. What is the nature of this peak?
Each momentum interval II, III, and  IV (Fig.~4) is related to the dominance of quasi-elastic scattering
 $^9$Be({\it p,~p$'$}~NC)$X$ ($X$ is a residual nucleus of the reaction)
 on a certain  low-nucleon correlation (NC). These reactiions are: $^9$Be($p,~p'$~$^2$H)$^7$Li for the interval II;
 $^9$Be($p,~p'$~$^3$He)$^6$He and $^9$Be($p,~p'$~$^3$H)$^6$Li for the interval III;
 $^9$Be($p,~p'$~$^4$He)$^5$He for the interval IV. A high-momentum region,  
 that begins close to the end of  the IV range (Figs.~3,~4) corresponds mainly to quasi-elastic 
 scattering on the residual nuclei $X$ from the reactions   $^9$Be({\it p,~p$'$}~NC)$X$ noted above:
 $^9$Be($p,~p'$~$^6$Li)$^3$H,  $^9$Be($p,~p'$~$^6$He)$^3$He,  $^9$Be($p,~p'$~$^7$Li)$^2$H, and
 $^9$Be($p,~p'$~$^5$He)$^4$He. This overlap of the momentum intervals 
 for the reaction $^9$Be($p,~p'$~$^4$He)$^5$He (interval IV) and, mainly, for the reactions 
 $^9$Be($p,~p'$~$^6$Li)$^3$H and  $^9$Be($p,~p'$~$^6$He)$^3$He with the light nucleus, $^9$Be,  is due to
  relatively close masses of the correlations and residual nuclei.
  
  In Fig.~3, there is a second rather wide peak in the momentum distribution with energy $\omega$ = $\omega_1$ = 9.7 MeV
transferred into the $^9$Be nucleus. The kinematic calculation shows that this peak possibly corresponds to quasi-elastic scattering on a $^8$Be - like nucleon cluster
in the momentum interval VIII.
The detection of this peak supports a theoretical model of the $^9$Be nucleus, within which the nucleus consists of a solid nucleon core 
 ($^8$Be - like nucleon cluster) and an external neutron weakly bound to this core \cite{Chavchanidze1951}. 
A value of the  measured polarization of the secondary protons produced at the scattering angle
$\Theta$~=~21$^\circ$  in the momentum interval VIII is given in Fig.~4.

 The measured momentum distribution  of the secondary protons
produced  in the reaction $^{9}$Be($p, p'$)$X$   at the scattering angle  $\Theta$~=~24.5$^\circ$ is shown in Fig.~5.
\clearpage
This distribution (Fig.~5) likes to that shown in Fig.~3. There are also  two wide peaks corresponding to the transfered energy  $\omega$ = $\omega_2$ = 35 MeV
and  $\omega$ = $\omega_1$ = 10.6 MeV. The peak at the $\omega$ = $\omega_1$  is possibly related to
quasi-elastic scattering on a $^8$Be - like nucleon cluster.

In Fig.~6,
the measured momentum distribution  of the secondary protons
produced  in the reaction with carbon nucleus $^{12}$C($p,~p'$)$X$   at the scattering angle  $\Theta$~=~21$^\circ$ is shown.
A stepwise similar drop in the momentum distribution is observed, which corresponds possibly to quasi-elastic scattering on
a nucleon cluster (NCL) inside the $^{12}$C nucleus ($^9$Be,  $^9$B,  $^{10}$B, and  $^8$Be)  in a reaction $^{12}$C($p$, $p'$~NCL)NC. 
Where NC is corresponding few-nucleon correlation ($^3$He, $^3$H, $^2$H, and $^4$He). This momentum distribution at the $K$ momentum
greater than 1640 MeV/$c$ is shown in Fig.~7. According to kinematic calculations, a rather narrow peak at  $K$ = 1658 MeV/$c$  (at the $\omega_1$  = 14.9 MeV)
 corresponds likely to  quasi-elastic scattering $^{12}$C({\it p,~p$'$}~$^8$Be)$^4$He on a $^8$Be - like nucleon cluster inside the carbon nucleus. 
 A value of the  measured polarization of the secondary protons in the quasi-elastic scattering on this cluster  is given in Fig.~8.
\begin{figure}
\centering\epsfig{file=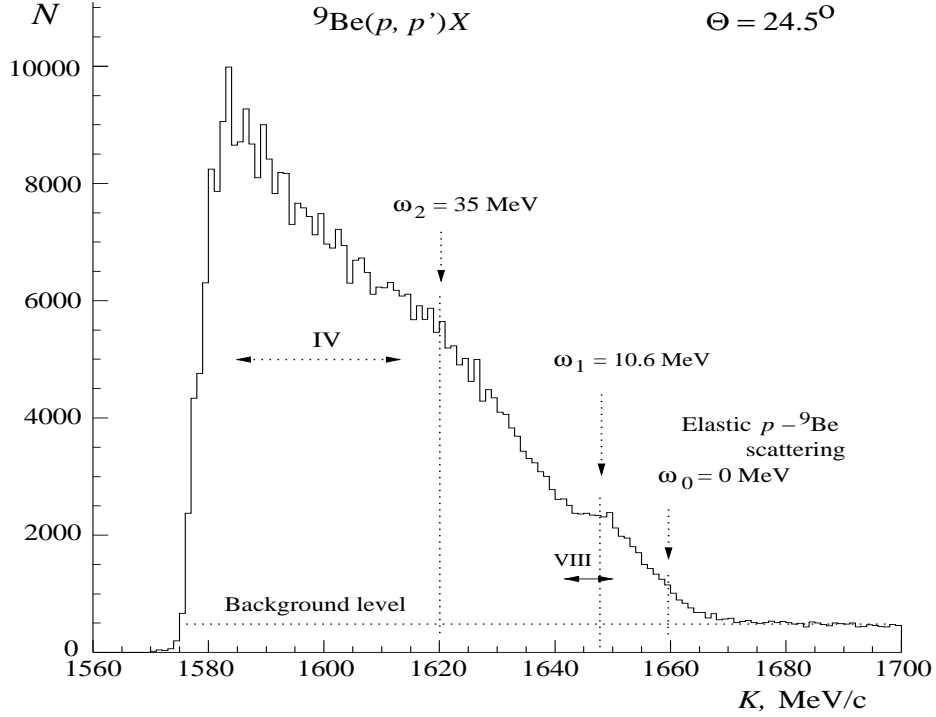,width=.85\textwidth,height=95mm}
\caption{\small   Momentum distribution of the protons scattered  at an angle $\Theta$~=~24.5$^\circ$ 
 in the inclusive reaction $^{9}$Be({\it p,~p$'$})$X$. $K$ is the 
secondary proton momentum. 
 A difference of the secondary proton energy calculated for the elastic proton scattering off the nucleus
 under investigation and measured in the experiment
 is signified by $\omega$. 
 The momentum interval VIII (IV) corresponds to  quasi-elastic scattering 
 on a $^{8}$Be - like (a $^{4}$He - like)  nucleon cluster.
 }
\end{figure}
%
\begin{figure} 
\centering\epsfig{file=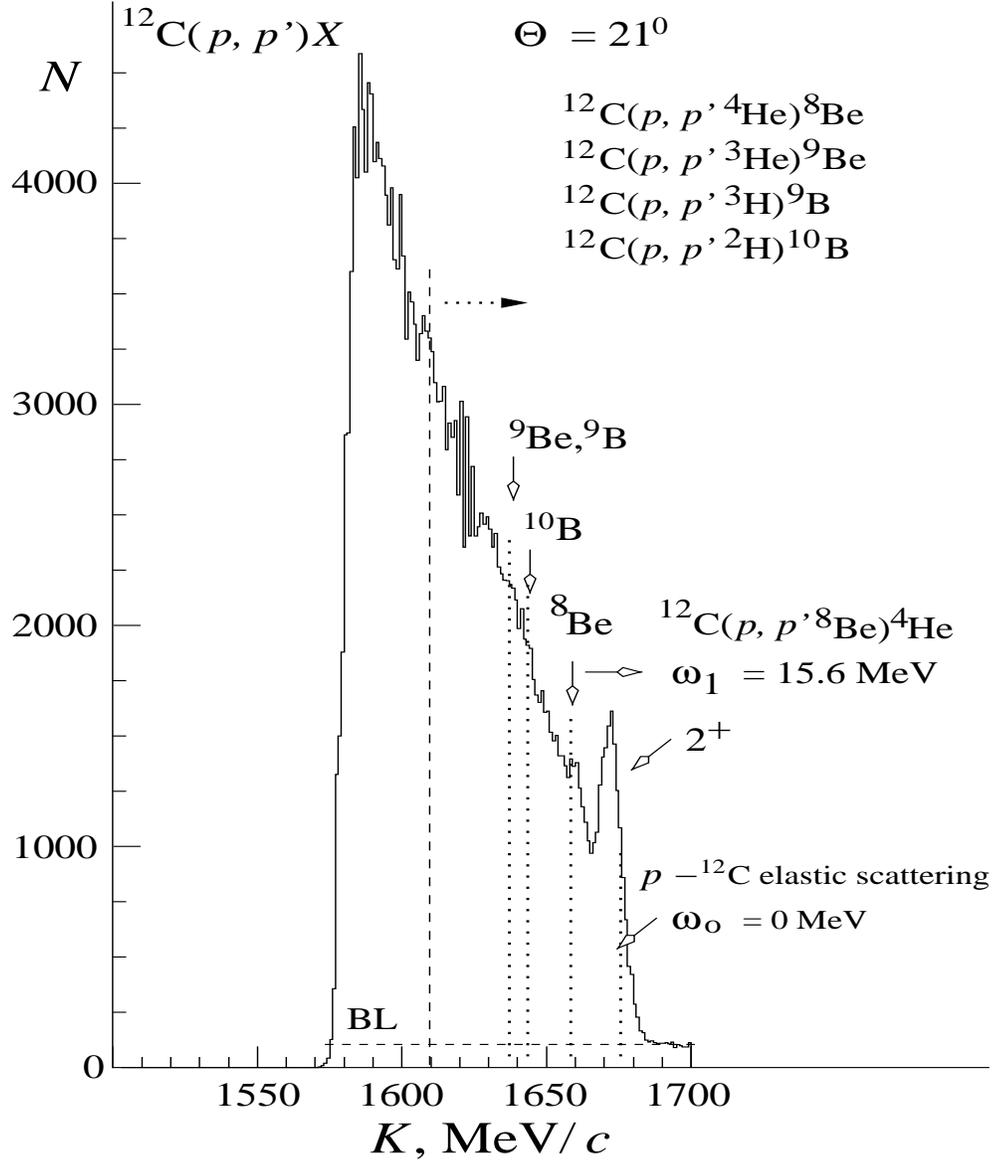,width=.90\textwidth,height=155mm}
\caption{\small   Momentum distribution of the protons scattered  at an angle $\Theta$~=~21$^\circ$ 
 in the inclusive  reaction $^{12}$C($p, p'$)$X$. $K$ is the
secondary proton momentum. 
Dashed vertical  line at the
$K$ = 1610 MeV/$c$ indicates the region of large $K$ , where effective registration of the secondary protons is carried out.
In the upper right corner of the figure, reactions are presented in which the proton scattering
 by few-nucleon correlations ($^2$H, $^3$H, $^3$He, and $^4$He) was studied.
 Vertical arrows point to
the calculated maxima of quasi-elastic peaks in scattering by residual nuclei ($^8$Be, $^9$Be, $^9$B, and $^{10}$B)  in the reactions noted above.
 A difference of the secondary proton energy calculated for the elastic proton scattering off the nucleus
 under investigation and measured in the experiment  is signified by $\omega$. Moreover $\omega$ = $\omega_0$  and $\omega$ = $\omega_1$
correspond to elastic scattering on the nucleus under study and  quasi-elastic scattering on a $^8$Be-like nucleon cluster inside the $^{12}$C nucleus.
BL means background level.
 }
\end{figure}
\clearpage

 The measured momentum distribution  of the secondary protons
produced  in the reaction with carbon nucleus $^{12}$C($p,~p'$)$X$   at the scattering angle  $\Theta$~=~24.5$^\circ$ is presented in Fig.~9.
This distribution  is similar to the momentum distribution at the scattering angle  $\Theta$~=~21$^\circ$ (Fig.~6).
A stepwise similar drop in the momentum distribution is also observed, which corresponds possibly to quasi-elastic scattering on
a nucleon cluster (NCL) inside the $^{12}$C nucleus ($^9$Be,  $^9$B,  $^{10}$B, and  $^8$Be)  in a reaction $^{12}$C($p$, $p'$~NCL)NC. 
Where NC is corresponding few-nucleon correlation ($^3$He, $^3$H, $^2$H, and $^4$He).
According to kinematic calculations, a peak at  the transferred energy  $\omega_1$  = 18.4 MeV to nucleus under investigation
 corresponds likely to  quasi-elastic scattering $^{12}$C($p$, $p'$~$^8$Be)$^4$He on a $^8$Be - like nucleon cluster inside the carbon nucleus. 
 
 When studying the ($p, p'$) reaction with $^{40}$Ca nucleus at a scattering angle of the secondary protons of 21$^{\circ}$,
  no structure was found in their momentum distribution,
  that could indicates proton scattering by a $^8$Be-like nucleon cluster inside this nucleus. However, at a scattering angle of 24.5$^{\circ}$ (Fig.~10), a bump
   in the momentum distribution is observed, that, according to kinematic calculations, can be associated 
   with scattering by a $^8$Be-like nucleon cluster inside the  $^{40}$Ca nucleus. In Fig.~11 
   corresponding to scattering at an angle of 21$^{\circ}$ on the Ca nucleus, an estimate of the polarization in quasi-elastic scattering on this cluster is given. 

\begin{figure}
\centering\epsfig{file=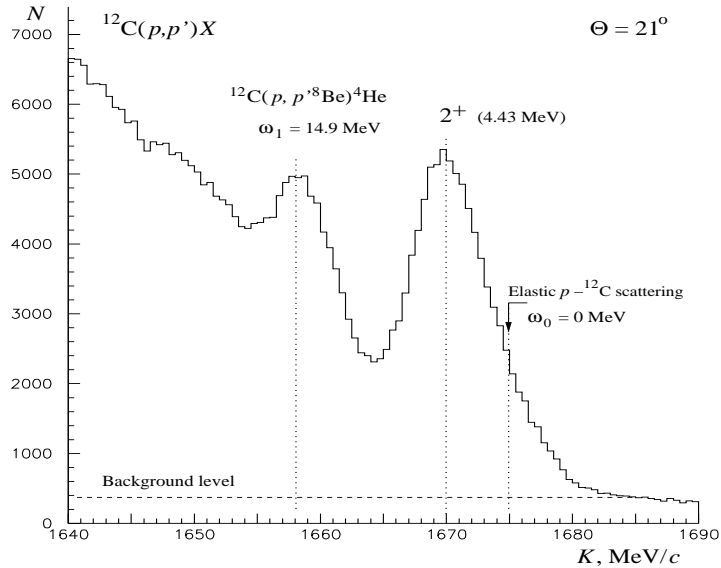,width=.65\textwidth,height=75mm}
\caption{\small   Momentum distribution of the protons scattered  at an angle $\Theta$~=~21$^\circ$ 
 in the inclusive reaction $^{12}$C($p, p'$)$X$. $K$ is the
secondary proton momentum. 
 A difference of the secondary proton energy calculated for the elastic proton scattering off the nucleus
 under investigation and measured in the experiment  is signified by $\omega$. Moreover $\omega$ = $\omega_0$  and $\omega$ = $\omega_1$
correspond to elastic scattering on the nucleus under study and  quasi-elastic scattering on a $^8$Be-like nucleon cluster inside the $^{12}$C nucleus.
 }
\end{figure}
%
\begin{figure}
\centering\epsfig{file=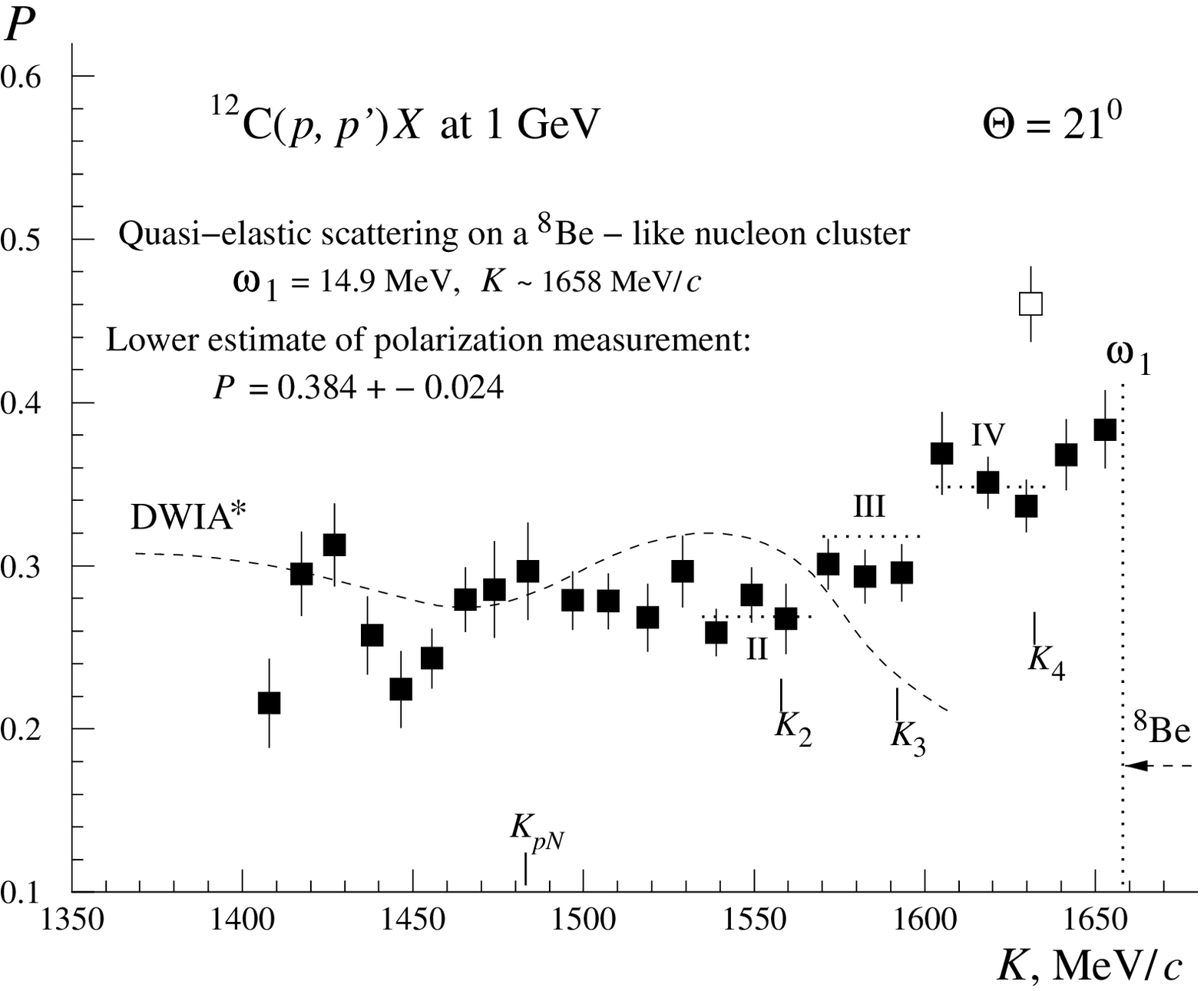,width=0.95\textwidth,height=120mm}
\caption{\small  Polarization $P$ of the protons scattered  at an angle $\Theta$~=~21$^\circ$
(black squares) in the inclusive reaction $^{12}$C($p, p'$)$X$ versus the
secondary proton momentum $K$ \cite{Miklukho2017, MiklukhaC2017}. 
The empty square corresponds to the polarization
in the elastic $p-^4$He scattering \cite{Miklukho2006}. The dotted lines cover
the $K$ intervals II, III, and IV corresponding to quasi-elastic scattering on two-nucleon ($^2$H), three-nucleon ($^3$He, $^3$H),
and four-nucleon ($^4$He) correlations, respectively. The calculated secondary proton momenta 
for the maxima of the quasi-elastic peaks in the $^{12}$C($p, p'$~NC)$X$ reaction with the corresponding nucleon correlation (NC) are designated as
 $K_2$, $K_3$, and $K_4$. 
 Momentum $K_{pN}$ corresponds to the maximum of the quasi-elastic peak in the proton scattering off nuclear nucleons.
  The dashed curve  presents the polarization calculated
in the framework of a spin-dependent Distorted Wave Impulse Approximation
 taking into account 
the relativistic distortion of the nucleon spinor in nuclear medium (DWIA*) \cite{Miklukho2017}.
 In this approach 
the proton scattering off the independent nuclear nucleons was taken into account only.
The energy $\omega_1$ = 14.9 MeV transferred to the $^{12}$C nucleus (Fig.~7)  
corresponds to quasi-elastic scattering on a $^8$Be-like nucleon cluster inside the nucleus.
}
\end{figure}
%
\begin{figure}
\centering\epsfig{file=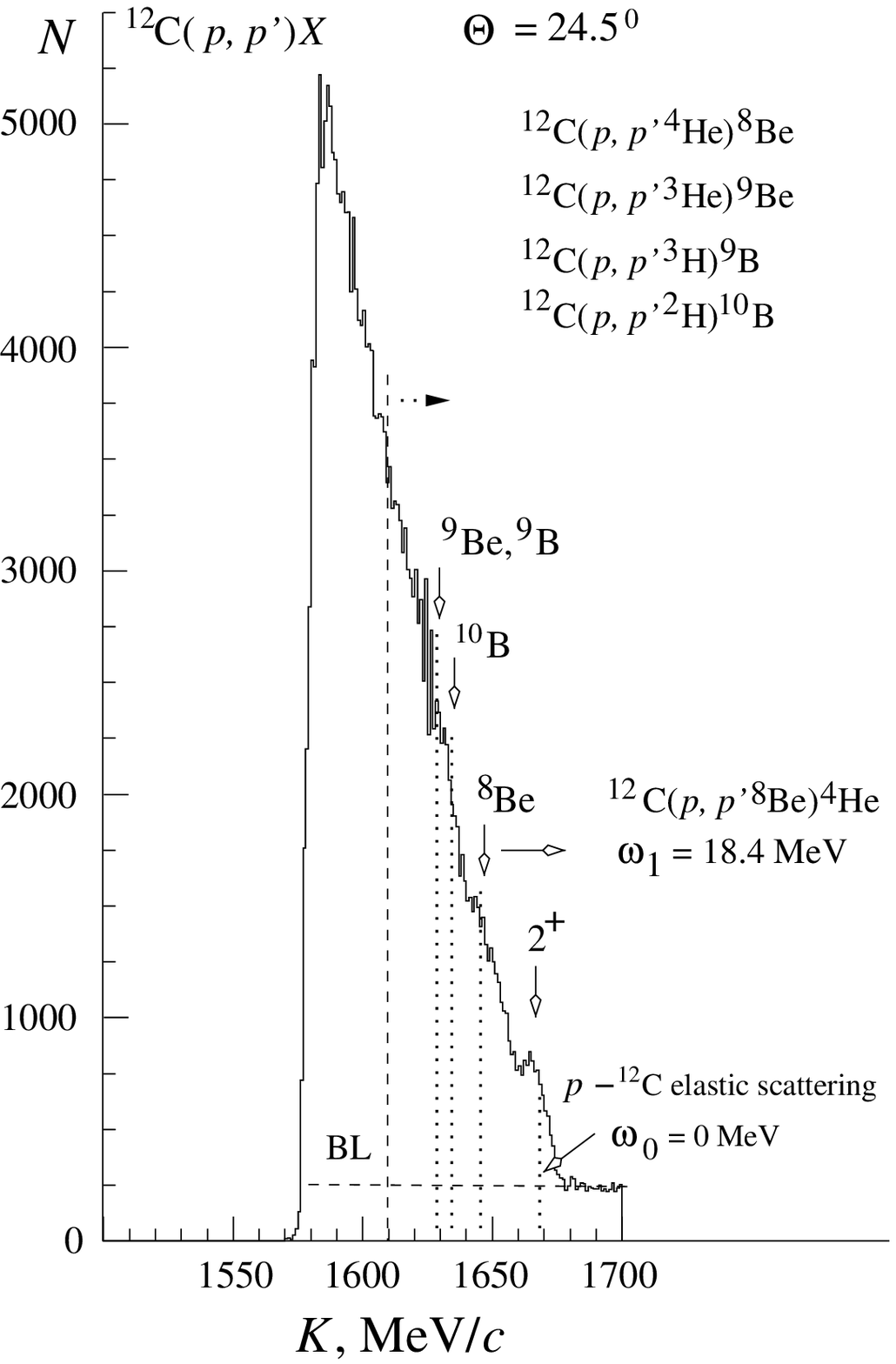,width=.90\textwidth,height=155mm}
\caption{\small   Momentum distribution of the protons scattered  at an angle $\Theta$~=~24.5$^\circ$ 
 in the inclusive reaction $^{12}$C($p, p'$)$X$. $K$ is the
secondary proton momentum. 
Dashed vertical  line at the
$K$ = 1610 MeV/$c$ indicates the region of large $K$, where effective registration of the secondary protons is carried out.
In the upper right corner of the figure, reactions are presented in which the proton scattering
 by few-nucleon correlations ($^2$H, $^3$H, $^3$He, and $^4$He) was studied.
 Vertical arrows point to
the calculated maxima of quasi-elastic peaks in scattering by residual nuclei ($^8$Be, $^9$Be, $^9$B, and $^{10}$B)  in the reactions noted above.
 A difference of the secondary proton energy calculated for the elastic proton scattering off the nucleus
 under investigation and measured in the experiment  is signified by $\omega$. Moreover $\omega$ = $\omega_0$  and $\omega$ = $\omega_1$
correspond to elastic scattering on the nucleus under study and  quasi-elastic scattering on a $^8$Be-like nucleon cluster inside the $^{12}$C nucleus.
BL means background level.
 }
\end{figure}
%
\begin{figure}
\centering\epsfig{file=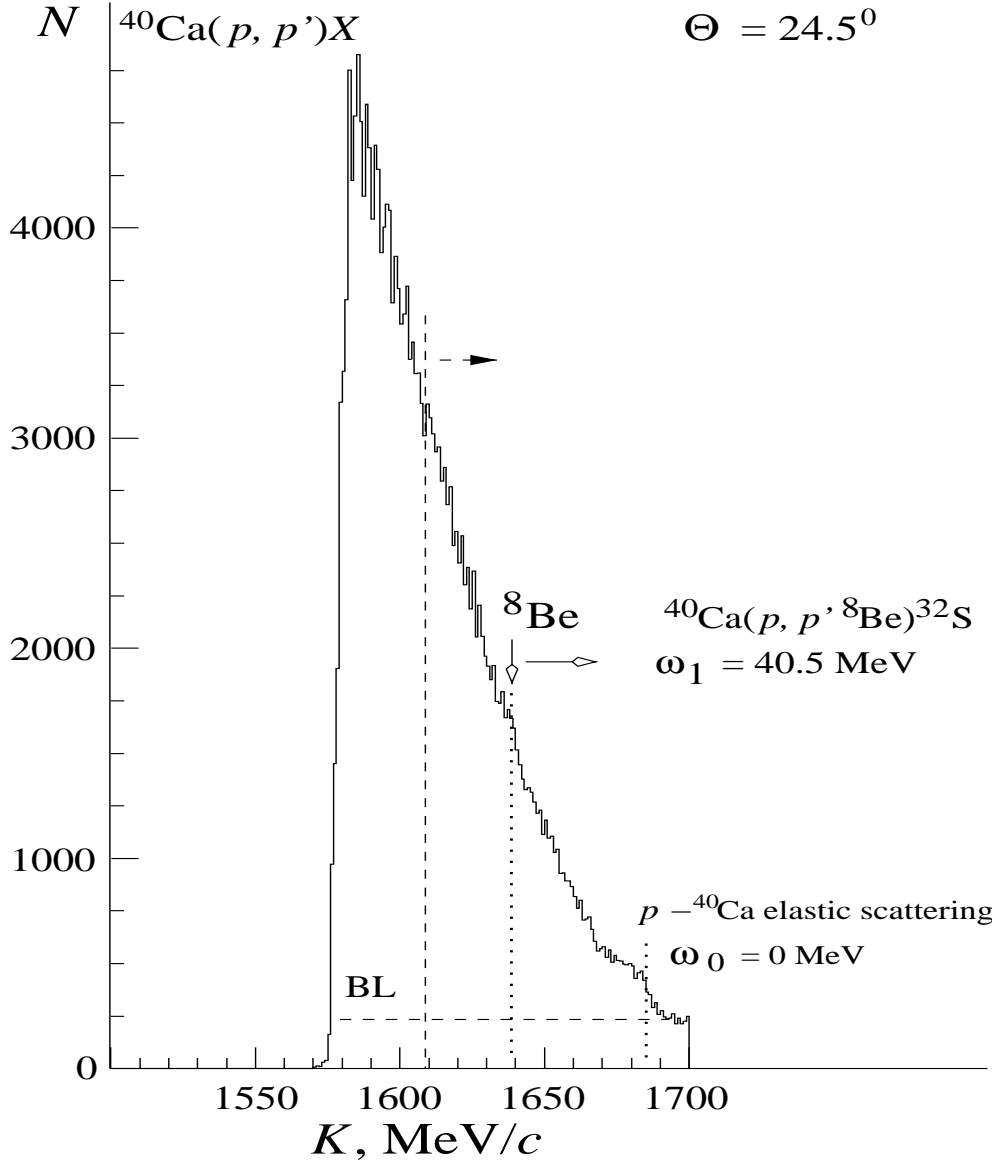,width=0.90\textwidth,height=155mm}
\caption{\small   Momentum distribution of the protons scattered  at an angle $\Theta$~=~24.5$^\circ$ 
 in the inclusive reaction $^{40}$Ca($p, p'$)$X$. $K$ is the
secondary proton momentum. 
Dashed vertical line at the
$K$ = 1610 MeV/$c$ indicates the region of large $K$, where effective registration of the secondary protons is carried out.
 Vertical arrow points to
the calculated maximum of quasi-elastic peak in the scattering $^{40}$Ca($p, p'$~$^8$Be)$^{32}$S
by a $^8$Be-like nucleon cluster inside the $^{40}$Ca nucleus. 
 A difference of the secondary proton energy calculated for the elastic proton scattering off the nucleus
 under investigation and measured in the experiment  is signified by $\omega$. Moreover $\omega$ = $\omega_0$  and $\omega$ = $\omega_1$
correspond to elastic scattering on the $^{40}$Ca  nucleus  and  quasi-elastic scattering on the $^8$Be-like nucleon cluster.
BL means  background level.
 }
\end{figure}

\begin{figure}
\centering\epsfig{file=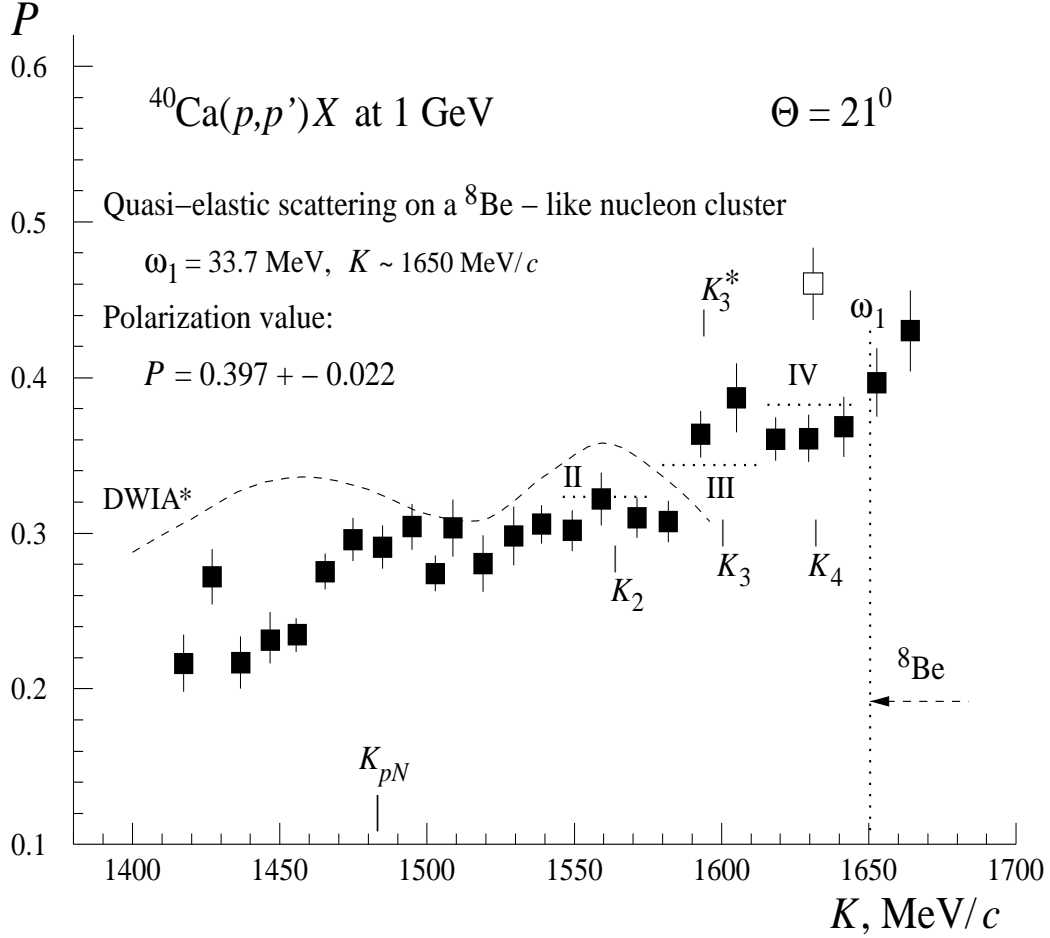,width=0.95\textwidth,height=125mm}
\caption{\small  Polarization $P$ of the protons scattered  at an angle $\Theta$~=~21$^\circ$
(black squares) in the inclusive reaction $^{40}$Ca($p, p'$)$X$ versus the
secondary proton momentum $K$ \cite{Miklukho2017}. 
The empty square corresponds to the polarization
in the elastic $p-^4$He scattering \cite{Miklukho2006}. The dotted lines cover
the $K$ intervals II, III, and IV corresponding to quasi-elastic scattering on two-nucleon ($^2$H), three-nucleon ($^3$He, $^3$H),
and four-nucleon ($^4$He) correlations, respectively. The calculated secondary proton momenta 
for the maxima of the quasi-elastic peaks in the $^{12}$C($p, p'$~NC)$X$ reaction with the corresponding nucleon correlation (NC) are designated as
 $K_2$, $K_3$ ($K_3^*$), and $K_4$. 
 Momentum $K_{pN}$ corresponds to the maximum of the quasi-elastic peak in the proton scattering off nuclear nucleons.
  The dashed curve  presents the polarization calculated
in the framework of a spin-dependent Distorted Wave Impulse Approximation
 taking into account 
the relativistic distortion of the nucleon spinor in nuclear medium (DWIA*) \cite{Miklukho2017}.
 In this approach 
the proton scattering off the independent nuclear nucleons was taken into account only.
The energy $\omega_1$ = 33.7 MeV transferred to the $^{40}$Ca nucleus  
corresponds to quasi-elastic scattering on a $^8$Be-like nucleon cluster inside the nucleus.
}
\end{figure}

\clearpage

\section{Summary}

A kinematic analysis of  momentum distributions of the secondary protons in an inclusive ({\it p, p$'$}) reaction with $^{9}$Be, $^{12}$C, and $^{40}$Ca nuclei    
 at an energy of 1 GeV and scattering angles of 21$^{\circ}$ and 24.5$^{\circ}$ is carried out.
 In a range of the high momenta close to the momentum  corresponding to the proton elastic scattering off the nucleus under study, the
 analysis indicates quasi-elastic scattering by a $^8$Be-like nucleon cluster inside the nucleus. This observation supports a theoretical model
 of $^{9}$Be nucleus, within which the nucleus consists of a solid nucleon core and an external  neutron weakly bound to
  this core.
  
 For the scattering angle $\Theta$~=~21$^\circ$, an estimate of the secondary proton polarization in quasi-elastic scattering on 
  a $^8$Be-like nucleon  cluster is given for nuclei under investigation.

   The authors are grateful to the PNPI 1~GeV proton accelerator staff
for the stable beam operation.\\
\\
\\

\end{document}